\documentclass[aps,prl,groupedaddress,reprint,showpacs, pdftex]{revtex4-1}
\usepackage{graphicx}
\usepackage{comment}
\usepackage{tabularx}
\usepackage{dcolumn}
\usepackage{bm}
\usepackage{amsmath}
\usepackage{amssymb}
\usepackage{txfonts}
\usepackage{url}
\usepackage{color}
\usepackage{multirow}
\newcommand{\x}{$\times$}
\newcommand{\tineut}{Ti$_{\rm i}^{\times}$}
\newcommand{\tiposi}{Ti$_{\rm i}^{\bullet\bullet\bullet\bullet}$}

\newcommand{\oneut}{V$_{\rm O}^{\times}$}
\newcommand{\oposi}{V$_{\rm O}^{\bullet\bullet}$}
\begin{document}
\title{
  Ti interstitial flows giving rutile TiO$_2$ reoxidation
  process enhanced in (001) surface
}
\author{Tom Ichibha}
\email[]{ichibha@icloud.com}
\affiliation{School of Information Science, JAIST,
  1-1 Asahidai, Nomi, Ishikawa, 923-1292, Japan}
\author{Anouar Benali}
\email[]{benali@anl.gov}
\affiliation{Computational Science Division, Argonne National
  Laboratory, 9700 Cass Avenue, Lemont, Illinois 60439, USA}
\author{Kenta Hongo}
\affiliation{Research Center for Advanced Computing
  Infrastructure, JAIST, 1-1 Asahidai, Nomi, Ishikawa 923-1292, Japan}
\affiliation{Center for Materials Research by Information Integration,
  Research and Services Division of Materials Data and Integrated System,
  National Institute for Materials Science, 1-2-1 Sengen,
  Tsukuba 305-0047, Japan}
\affiliation{PRESTO, Japan Science and Technology Agency, 4-1-8 Honcho,
  Kawaguchi-shi, Saitama 322-0012, Japan}
\affiliation{Computational Engineering Applications Unit, RIKEN,
  2-1 Hirosawa, Wako, Saitama 351-0198, Japan}
\author{Ryo Maezono}
\affiliation{School of Information Science, JAIST, 1-1 Asahidai,
  Nomi, Ishikawa, 923-1292, Japan}
\affiliation{Computational Engineering Applications Unit,
  RIKEN, 2-1 Hirosawa, Wako, Saitama 351-0198, Japan}
\date{\today}
\begin{abstract}
We revisited {\it ab initio} evaluations 
of the energy barriers along the possible 
diffusion paths of the defects in rutile TiO$_2$. 
By using a method carefully considering 
the cancellation of the self-interaction, 
Ti interstitials hopping along $c$-axis 
are identified as the major diffusion 
directing to [001] surface. 
The conclusion is contradicting to 
any of previous theoretical works, 
and the discrepancy 
is explained by the overestimation of 
the radius of defects due to the poor 
cancellations in the previous works.
The updated prediction here can explain the 
superior photocatalysis activity in 
[001] surface to [110]. 
\end{abstract}
\maketitle

TiO$_2$ is a representative transition metal oxide
with various applications such as white paints,
photovoltaic cells, and rechargeable batteries.
~\cite{2016ALV, 2017LON, 2012ABBa, 2012ABBb, 2018GHA} 
Its photo-catalysis ability is especially useful for
water splitting and anti-pollution/bacteria coating.~\cite{2014VER}
During the photo-catalysis reaction, O ions are easily detached
from the surface,~\cite{2007PAR}
and hence one may anticipate the depression of
the photo-catalysis ability.
Yet in reality, the surface gets O ions from the atmosphere,
and the photo-catalysis ability is maintained.~\cite{2007PAR}

\vspace{2mm}
One of the most useful properties is the 
reoxidization of rutile surface state even in a vacuum 
keeping its stoichiometry. 
The property is promising for such applications 
in space as a coating over the solar panels 
of spaceships keeping its performance of photo 
reactions.\cite{2007URA} 
The reoxidization in a vacuum is 
explained to be caused by the possible 
ionic flows of Ti interstitials (Ti$_\mathrm{i}$) 
and/or Oxygen vacancies (V$_\mathrm{O}$) from within the bulk
toward the surface compensating the 
stoichiometry kept unchanged.~\cite{2007PAR}
  However, a consensus on the diffusion process of point
  defects has yet to be established and controversy
  remains even for within a simple bulk structure.~\cite{2007IDD,2010ASA}

\vspace{2mm} 
Surveying over the controversy, 
the points to be clarified here would be  
summarized into two simple questions: 
(a) which defect (Ti$_\mathrm{i}$ or V$_\mathrm{O}$)
is the dominant, and 
(b) which diffusion path is dominating. 
An experiment of the reoxidization 
of the sputtered rutile TiO$_2$ (110) surface 
annealed in ultrahigh vacuum~\cite{1999HEN} 
reports a conclusion that Ti$_\mathrm{i}$ plays 
a major role in the process. 
This is also supported from {\it ab initio} 
studies using density functional theory (DFT),
~\cite{2007IDD, 2010ASA} 
predicting lower energy barriers for 
Ti$_\mathrm{i}$ than V$_\mathrm{O}$ diffused in any directions. 
Taking Ti$_\mathrm{i}$ being superior to V$_\mathrm{O}$, 
the controversy exists on which path 
gives faster diffusion, 
parallel ($c_\parallel$) or 
perpendicular ($c_\perp$) to 
the $c$-direction [parallel to the Ti-chain in the crystal].
While two old experiments~\cite{1965HUN,1985HOS}
report contradicting conclusions to each other,
both of the previous DFT works~\cite{2007IDD,2010ASA}
support $c_\perp$ as the major diffusion process. 

\vspace{2mm}
One of the major origin of the energy barrier 
required for a defect to move beyond 
is the interaction between the surrounding atoms. 
It is therefore sensitive to how the electronic 
distribution of a defect spreads to contact 
with the neighboring atoms. 
Here we remind that such a spreading 
is poorly estimated by the conventional type 
of DFT using LDA or GGA type 
exchange-correlation (XC) functionals. 
In these XCs, the cancellation of the self-interaction 
is incomplete, leading to a spurious delocalization 
of the charge distribution.~\cite{2009GOR,2018BAO} 
This incompleteness can be corrected 
by using 'DFT$+U$' scheme for the 
self-interaction.~\cite{2014HIM}
The method mainly enhances the exchange
part of the conventional XC.
Yet, the balance between the exchange 
and the correlation in XC is delicate~\cite{2006HAR}
and should be preserved to get more reliable predictions.
Diffusion Monte Carlo (DMC) method is the 
most reliable method in the sense that
such a delicate balance is automatically 
fulfilled in a numerical implementation of the variational 
principle.~\cite{2001FOU}
The method was successfully applied to the present
TiO$_2$ system in previous works~\cite{2016LUO,2017TRA,2012ABBb,2018GHA}. 

\vspace{2mm}
In this work, we hence revisit the evaluation 
of the energy barrier for defects applying DMC. 
We confirmed that Ti$_\mathrm{i}$ is the dominant defect to diffuse,
contributing to the reoxidation process with an energy barrier lower
than that for V$_\mathrm{O}$, being consistent with previous DFT works. 
~\cite{2007IDD,2010ASA}
A striking finding we made is that 
the previous DFT prediction 
supporting $c_\perp$ is reverted into 
$c_\parallel$ when the cancellation of
the self-interaction is considered by using '+$U$' or DMC.
The results support a better reoxidation activity on (001) surface,
consistent with experiments~\cite{1998LOW,1998MOR} reporting that
the said surface has almost the highest photo-activity.

\begin{figure}[htbp]
  \begin{center}
    \includegraphics[width=\hsize]{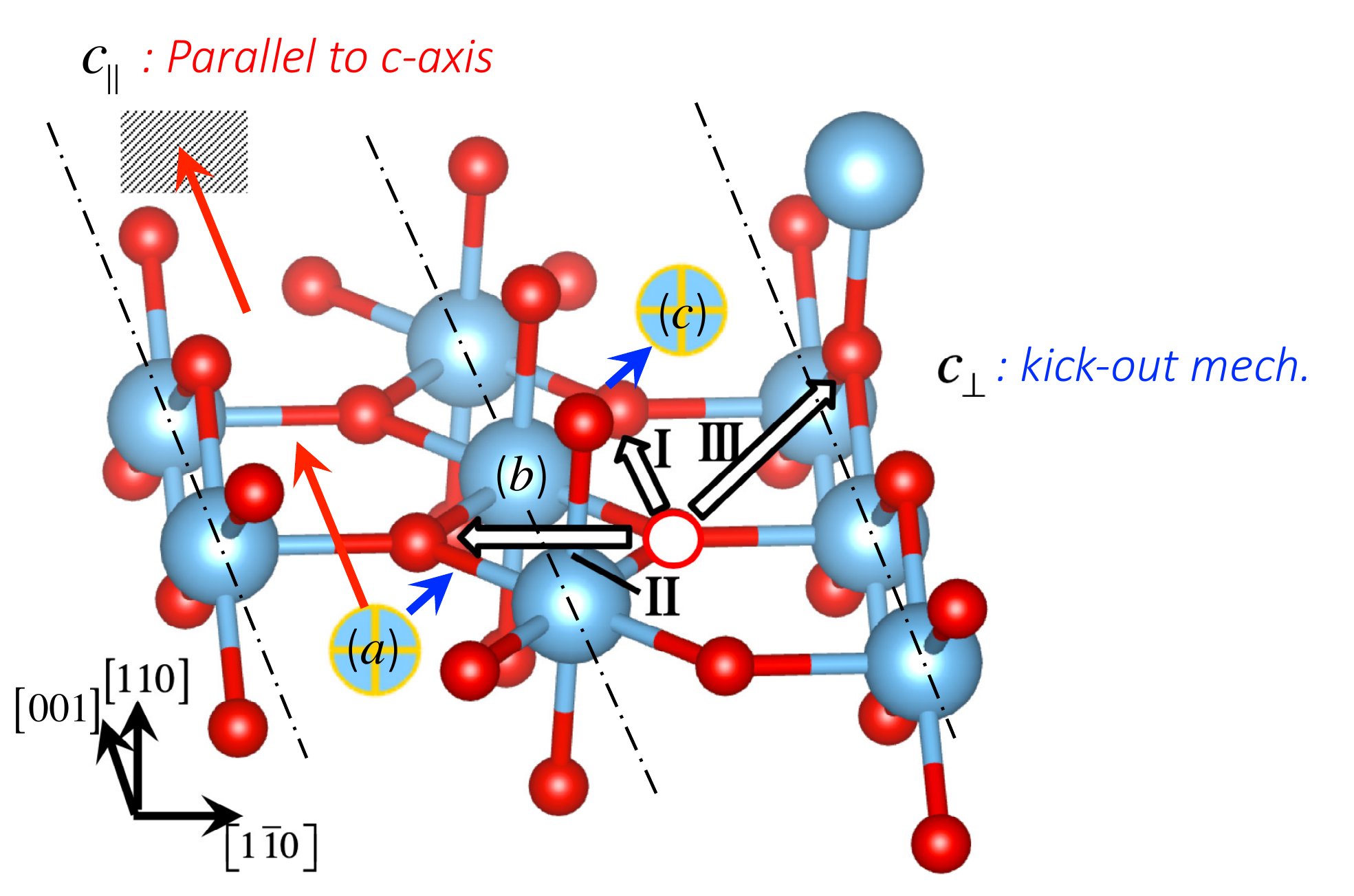}
  \end{center}
  \caption{[fig.ti\_diff]   
    Five possible paths for defect diffusions 
    of Ti$_\mathrm{i}$ (blue and red arrows) and 
    V$_\mathrm{O}$ (white arrows) in bulk rutile TiO$_2$. 
    The large blue balls are Ti ions and
    the red small balls are oxygen ions.
    Ti atoms are located along the 
    $c$-axis ([001]-direction). 
    In $c_\perp$ diffusion (blue arrow), 
    a Ti$_\mathrm{i}$ kicks a Ti on the axis out 
    to make another Ti$_\mathrm{i}$ in opposite side 
    (kick-out diffusion\cite{1985SAS}), 
    directing along [100] or [010] axis. 
    The diffusion along the path $c_\parallel$ 
    (red arrow) directs toward [001] surface 
    as shown by a hatched square.
  }
  \label{fig.ti_diff}
\end{figure}

\vspace{2mm}
The rutile structure of TiO$_2$ is shown
in Fig.~\ref{fig.ti_diff}. 
It consists of Ti chains along the $c$-axis.
Ti positions along the axis are shifted  
by 1/2 period between the neighboring chains. 
Ti$_\mathrm{i}$ is formed in the middle of Ti chains 
as shown in Fig.~\ref{fig.ti_diff},~\cite{2007IDD} 
for which two possible diffusion paths 
($c_\parallel$ and $c_\perp$)
are of interest.~\cite{2007IDD}
The hopping along $c_\perp$ is described as  
the 'kick-out mechanism'.~\cite{1985SAS} 
For V$_\mathrm{O}$, three paths, I-III in Fig.~\ref{fig.ti_diff}, 
are considered.~\cite{2007IDD} 
We evaluated barrier energies along these five 
paths for fully positively charged 
defects~({\tiposi}, {\oposi}), 
as summarized in Table \ref{tab.ti_barrier}. 
Previous theoretical works~\cite{2012LEE,2007IDD}
predict only the possibility of getting
{\tineut},{\tiposi},{\oneut},{\oposi} 
depending on the Fermi level. 
Experimentally, the charged defects are 
confirmed to be realized in surface,~\cite{2015NOW}
and hence we took {\tiposi} and {\oposi} 
as the defects to be investigated. 
The results by the neutral defect, 
{\tineut}, are also shown in 
Table \ref{tab.ti_barrier}, which 
are referred only when we make 
further discussions. 
The descriptions henceforth are 
therefore about the {\tiposi} and {\oposi} 
unless noted otherwise.

\vspace{2mm}
We made a simulation cell by putting a point defect 
into a 2{\x}2{\x}3 supercell of the ideal rutile 
TiO$_2$ unit cell. 
We optimized the crystal structures 
at the {\it edge} and the {\it saddle} 
points of the states along the diffusion paths 
using the PAW-DFT method implemented on VASP.\cite{1996KRE} 
The optimizations are made to relax 
internal atomic positions within a cell under 
the fixed lattice constants at experimental 
values.\cite{1998ISA} 
The energy cutoff is 700~eV and the spacing of 
the $k$-mesh sampling is denser than 0.50~$\AA^{-1}$. 
Atomic positions are relaxed until the forces on 
any ions are suppressed less than 0.01~eV/$\AA$. 
The structures at the saddle states are 
determined by the climbing nudged
elastic band (c-NEB) method.\cite{2000HEN} 
  A diffusion path is expressed with 5 or 15
  intermediate states between the edge states.
  Since one of the states must be converged
  to be the saddle state in c-NEB,~\cite{2000HEN}
  the number of states does not affect
  the barrier energy prediction
  but affect the convergence of the relaxation.

\vspace{2mm}
We applied DMC to evaluate the energies at the edge and
saddle structures using QMCPACK.~\cite{qmcpack1}
We used Slater-Jastrow type trial wave 
functions.~\cite{2001FOU,2009MAE}
Orbital functions used in the Slater determinant 
are generated by LDA+$U$ method implemented in 
Quantum Espresso.~\cite{2009GIA}
We used a Hubbard correction value of $U$=4.86 eV from
a previous work,~\cite{2016LUO}
giving the best accessible nodal surface within this formalism,
guaranteeing the lowest energy for TiO$_2$ from the variational principle.
Core electrons in both Ti and O atoms were described by the use of
a hard norm-conserving pseudopotentials developed to reproduce
accurately all electrons results with the context of many-body theory
and as described in previous works~\cite{2016LUO}.
The orbitals are generated with a 300~Ry energy cutoff
and the thermodynamic limit is reached with a 2$\times$2$\times$2
$k$-mesh size.
The Jastrow factor consists of one, two, and three body terms 
amounting to 144 variational parameters in total, 
which are optimized by variational Monte Carlo 
calculations.~\cite{2001FOU,2009MAE} 
The parameters are optimized by 
the scheme to minimize a hybridization 
of energy and variance in 7:3. 
Twist averaging over the boundary conditions 
are taken into account with 
2$\times$2$\times$2 grid.~\cite{2001LIN} 
We estimated a time-step bias by a linear extrapolation 
of the energies obtained at two time steps, 
$dt=0.020$ and $0.005$~a.u.$^{-1}$. 
It is confirmed that the time-step bias is 
proportional to $dt$ in a range of 
$dt$ $<$ 0.020~a.u.$^{-1}$. 
We set a target population of walkers to be 4,000. 
Practically this size of target population is
large enough to suppress a population control error. 

\vspace{2mm}
Table~\ref{tab.ti_barrier} summarizes the 
results of the barrier energies along each path.
Looking at the lowest barrier-energies (shown in bold),
all methods, consistent with each other, predict
Ti$_\mathrm{i}$ as the preferred diffusion carrier.
The striking difference is found
between our current result and the previous ones
regarding Ti$_\mathrm{i}$ preferred diffusion path.
Updated predictions by LDA+$U$ and DMC 
supports $c_\parallel$ as the
dominant flow, directing towards the (001) surface
while $c_\perp$ directing towards the (100) or (010) surface.
The prediction here may explain the experimental observation 
of the photoactivity being enhanced at (001) surface
compared to the (110) surface.~\cite{1998LOW,1998MOR}
We note that LDA$+U$ and DMC give 
different predictions about the fastest
diffusion path for V$_\mathrm{O}$.
Our final DMC prediction gives path II as the fastest path 
for oxygen vacancy diffusion (V$_\mathrm{O}$). 
However, path II alone cannot produce any diffusion flows 
because sites in this path are disconnected from each other. 
For V$_\mathrm{O}$s to diffuse globally in the bulk 
a combination of path I /III with path II is needed, 
otherwise V$_\mathrm{O}$s will be constrained to 
the isolated sites in path II.
\begin{table}
  \begin{center}
    \caption{\label{tab.ti_barrier}
    Barrier energies of Ti$_\mathrm{i}$ ($c_\parallel$ and $c_\perp$) 
    and V$_\mathrm{O}$ (I, II, and III) paths evaluated by 
    various methods, including previous works.
    \cite{2007IDD, 2010ASA} 
    All the predictions are made for fully positively 
    charged defects~({\tiposi}, {\oposi}),
    except 'DMC ({\tineut}' (neutral) which is shown
    for a reference in discussions.
    The geometries to evaluate the barrier 
    are optimized each to neutral and charged states, 
    independently. 
    }
    \begin{tabular}{c|cc|ccc}
      & \multicolumn{2}{c}{Ti$_\mathrm{i}$} & \multicolumn{3}{c}{V$_\mathrm{O}$} \\
      & $c_\parallel$ & $c_\perp$ & I & II & III \\
      \hline
      GGA-PW91\cite{2007IDD} & 0.37   & {\bf 0.225}  & 1.77 & \underline{0.69} & 1.1 \\
      GGA-PW91\cite{2010ASA} & 0.31   & {\bf 0.23}   & -- & -- & -- \\
      LDA+$U$                & {\bf 0.54}   & 0.90   & 2.42 & 1.60 & \underline{1.36} \\
      DMC                    & {\bf 0.4(1)} & 0.9(1) & 2.0(1) & \underline{0.9(2)} & 1.7(1) \\
      \hline
      DMC ({\tineut})        & 2.6(4) & 1.6(1) & -- & -- & -- \\ 
      \hline
    \end{tabular}
  \end{center}
\end{table}

\vspace{2mm}
When compared to our DMC results, previous GGA-DFT calculations 
show a significant underestimation of barrier energies. 
Even using ``the same fixed geometry relaxed with DFT+$U$'' in GGA and DFT+$U$ 
calculations, the trend of underestimation is confirmed.
This can be attributed to that GGA generally underestimates 
a cohesive energy~\cite{2019ARR}, since a defect is more weakly
combined with the surrounding ions than reality, making its hopping easier.

\vspace{2mm}
As can be seen in Table~\ref{tab.ti_barrier}, evaluating
the diffusion path of the neutral defect 'DMC ({\tineut})',
the most favorable diffusion path is $c_\perp$, opposite to
what is found for a charged defect.
This might be a clue to understanding 
why the present result is contradicting
to the previous DFT works, 
as well as to understanding 
the contradiction in the earlier 
experiments:~\cite{1965HUN,1985HOS} 
One of the dominant factor to determine 
the preferred diffusion
path could be the ionic radius of the defects,
which is reduced when they are positively charged 
to reduce accompanying electrons. 
The sensitive dependence on the choice of 
XC potentials in Table~\ref{tab.ti_barrier} 
could support this, because the 
estimation of the radius is known to be 
sensitive to how the self-interaction 
is carefully treated.~\cite{2018BAO} 
Poor treatments are expected to give 
a spurious delocalization of distribution
leading to a larger radius.~\cite{2009GOR} 
The Hubbard '$+U$' correction is introduced 
to correct this, and hence corrects 
the radius smaller. 
Previous GGAs are therefore suspected 
to give overestimations of the radius, 
namely '{\it spuriously less positively 
charged defects}'.~\cite{2009GOR} 
A Bader analysis using a scheme described in ref.\cite{2009TAN} 
actually estimates the volumes of defects as 
($\rho$,$V$)=(2.206, 6.765) for the charged state (+4)
while ($\rho$,$V$)=(1.773, 7.690) for the neutral state, 
where $\rho$ and $V$ denote the amount of accompanying 
charge and the volume of a defect (in \AA$^3$), 
respectively.

\vspace{2mm}
  An earlier experiment~\cite{1985HOS} supporting $c_\perp$
  as the preferred path was actually performed
  at high temperatures raging from 1000 to 1500~K.
  It is shown through simulation~\cite{2018SHA} that
  the electronic distribution in the valence region is expanded
  with high temperatures.
  The high temperature experiment suggests a less positively
  charged defect favoring the $c_\perp$ path.
  This behavior is confirmed by our DMC ({\tineut}) calculation
  on a neutral defect (see table~\ref{tab.ti_barrier}).

\vspace{2mm}
In conclusion, 
we performed {\it ab initio} evaluations of 
the energy barriers for defects of Ti interstitials 
and Oxygen vacancies using LDA$+U$ and DMC methods. 
Ti interstitials diffusing along the Ti-chains 
($c$-axis) are predicted to give the lowest 
energy barrier, being the most likely origin
of the atomic flow toward [001] 
surface supporting the surface reoxidizations. 
The result is consistent with the 
photocatalysis activity in [001] surface 
being superior to [110] as experimentally observed.~\cite{1998LOW,1998MOR} 
The prediction is found to be sensitive to how 
carefully the cancellation of self-interactions 
is taken into account, not reproduced by the 
conventional DFT with non-hybrid XC functionals.~\cite{2007IDD,2010ASA} 
The cancellation critically changes the radius 
of the defects interacting surrounding atoms,
which was overestimated by the previous DFT works.~\cite{2007IDD,2010ASA}

\section{Acknowledgments}
  An award of computer time was provided by the Innovative and
  Novel Computational Impact on Theory and Experiment (INCITE)
  program. This research used resources of the Argonne Leadership
  Computing Facility, which is a DOE Office of Science User
  Facility supported under Contract No. DE-AC02-06CH11357,
  and the Research Center for Advanced Computing Infrastructure
  (RCACI) at JAIST.
T.I. is grateful for financial suport from Grant-in-Aid
for JSPS Research Fellow (18J12653).
  A.B. is supported by the U.S. Department of Energy, Office of
  Science, Basic Energy Sciences, Materials Sciences and Engineering
  Division, as part of the Computational Materials Sciences Program
  and Center for Predictive Simulation of Functional Materials.
K.H. is grateful for financial support from a KAKENHI grant
(JP17K17762),
a Grant-in-Aid for Scientific Research on Innovative
Areas ``Mixed Anion'' project (JP16H06439) from MEXT, 
PRESTO (JPMJPR16NA) and the Materials research by
Information Integration Initiative (MI$^2$I) project 
of the Support Program for Starting Up Innovation Hub
from Japan Science and Technology Agency (JST). 
R.M. is grateful for financial supports from 
MEXT-KAKENHI (19H04692 and 16KK0097), 
from Toyota Motor Corporation, from I-O DATA Foundation, 
from the Air Force Office of Scientific Research 
(AFOSR-AOARD/FA2386-17-1-4049;FA2386-19-1-4015), 
and from JSPS Bilateral Joint Projects (with India DST). 
R.M. and K.H. are also grateful to financial supports
from MEXT-FLAGSHIP2020 (hp190169 and hp190167). 

\bibliography{references}
\end{document}